\documentclass[modern]{aastex631}

\usepackage{graphicx}
\usepackage{color}

\begin{document}

\title{What I Wish I had Known When I Began Building Astronomical Instruments}

\author[0000-0002-1311-4942]{Daniel Fabricant}
\email{dfabricant@cfa.harvard.edu}\affiliation{Center for Astrophysics | Harvard and Smithsonian\\
60 Garden Street\\
Cambridge, MA 02138, USA}

\begin{abstract}
This paper describes lessons learned over a long career building astronomical instruments.  Although these lessons are based on one person's experiences, they were learned while working on major and minor instrument projects with many scientific and engineering colleagues.  When I interact with colleagues entering the field, I am reminded that the``obvious" approaches were in fact learned over many years by observing others develop instruments, by receiving good advice, and by making mistakes.  I hope to help others avoid making all of the same mistakes.

\end{abstract}


\section{Introduction} \label{sec:intro}

This paper is a high level summary of habits of thought and approaches that I have found useful in leading the development of instruments for ground-based telescopes. Most should apply more generally to other types of projects.  When working with graduate students, postdoctoral fellows, and more senior scientists starting an instrument project, I often discuss these ideas.  This paper aims at greater completeness and clearer organization than is possible in a typical conversation.  Instruments are built in a variety of environments, so flexibility and a willingness to adapt are essential.   

\clearpage

\section{Personal Skills to Develop}

\subsection{Understand the Scientific Mission}

Instrumentalists are fond of saying that new instruments open new areas of research, but all new areas of research are not equally productive.  Filling empty niches and using new technology is not a guarantee of success.  Developing instruments that address a significant scientific question might make use of recent technical advances, but the scientific mission is the first consideration.  A first-rate instrumentalist should have enough scientific background to recognize first-rate science and to evaluate the strength of the scientific team defining the core scientific mission.

\subsection{Humble Curiosity} \label{Humility}

Humility and curiosity are important characteristics for instrument project leaders. It is possible that your new instrument breaks completely new ground and that you are uniquely qualified to develop this instrument.   More likely, a substantial literature and knowledgeable experts could save you a great deal of time if you took the time to explore.  Whatever your feelings about AI and large language models, these tools have made looking for information much easier.  Use them, read, and learn.  Contact experts that you come across in your reading. They may not always answer your queries, but many will if the answer is not easily available in the literature. Armed with the appropriate background you will need to think and possibly develop new approaches, but you can avoid reinventing the wheel. There is no virtue in reinvention that ignores generations of knowledge.

If you are lucky enough to be working at an institution that employs experienced scientists and engineers, talk to them.  Collectively, their knowledge greatly leverages your search for information.  Few things animate colleagues more than the opportunity to teach.  Rather than resenting the interruption, most colleagues are pleased to help.

\subsection{Connect Your Knowledge to a Bigger Picture}

Integrating bits of information and new approaches into a bigger picture is an important habit that validates new pieces of information and allows you to later more easily use that information.  We don't learn an entire new field at once, rather we slowly gain a working knowledge in pieces that we connect.  The pieces we learn excite our curiosity and motivate us to learn more.  Curiosity about something specific pulls us into new areas.

\subsection{Systems Engineering}

Modern instrumentalists will be familiar with the systems engineering process: developing and tracking instrument requirements from a list of scientific requirements.  The formality of the systems engineering process by necessity depends on the project scale.  NASA-scale projects are controlled by system engineers, but your smaller project may not have an available team of systems engineers familiar with requirements tracking software.  You may need to be your own systems engineer, and keep track of key instrument requirements yourself.  To play this role you must understand and value the scientific mission, and help the scientific users to understand the tradeoffs between the scientific mission and the instrument design.  Even in a larger project where system engineers are engaged, you will be the last defense against system engineering failures if you really understand the scientific mission.

Systems engineering, as implemented in most projects, cannot correct for areas of concern missed in the initial analysis, weak engineering expertise, or the absence of invested, knowledgeable leadership. The leadership should encourage frequent discussions between end users and the project team to correct oversights and miscommunications.

\subsection{Take Responsibility for Failures}

Real leadership requires accepting responsibility even when a poorly executed task was delegated to someone else.  There are many examples of those in authority taking credit for successes and blaming failures on others.  Shirked responsibility is demoralizing and long remembered.  Even successful instrument programs encounter failures and disappointments along the way so accept these and move forward.

\section{Instruments are Built by a Team of People}

\subsection{Organize Your Team First}

Although the focus in your education is on one person (you) successful instruments are designed, built, and tested by a team.  No one will be impressed if your individual work is brilliant but the team effort is a failure.  Getting the tasks done that you are personally contributing is necessary, but your first responsibility is to make sure that all team members know what is expected of them and have the tools to do their job.  Before you become absorbed by an interesting problem, make sure that all team members are working effectively on their problems.

\subsection{Engineers Are Not Interchangeable}

We sometimes assume that a degree in engineering fields comes with a guarantee of particular knowledge or experience.  This is not the case, so it is up to the consumer of engineering services to learn what can be expected when a new engineer is brought into a project.  Speak to colleagues who have previously worked with the engineer.  Set up short term progress milestones with the new engineer to provide feedback.  No progress that can be presented is no progress, so monitor the work carefully, particularly at the beginning.

I learned this lesson the hard way when I started working on the conversion of the MMT to use a 6.5m primary mirror.  My goal was to develop the optics that would give the converted MMT a 1$^{\circ}$ diameter field of view instead of the narrow field originally planned. The first order of business was to design the 1.8m diameter secondary mirror and its support and the wide-field corrector optics mount.  At the time, the main emphasis of our Central Engineering group was developing the Chandra Observatory and ground-based projects were considered secondary.  I was assigned engineers who were not considered essential for the main mission because they were temporary hires or untested.  I worked with two inexperienced and misguided engineers who produced bad results very slowly. When I discussed their  work with colleagues working on the MMT conversion, my concerns were reinforced.  I complained to engineering management, and in a stroke of luck was assigned Robert (Bob) Fata.  Although Bob was also new, he turned out to have considerable experience at Itek designing large lens and mirror mounts and performing structural analysis.  Bob showed himself to be as serious about solving engineering challenges as I am about building instruments, and we worked together for 30 years until his retirement.

\subsection{How Engineers Approach Design}

Engineers are trained to work from a list of requirements.  They are not always comfortable judging when it would be desirable to do better than just meeting your minimum requirements.  The engineers on your team may split their time between multiple projects and just meeting the minimum requirements may help them deal with time pressure. Nonetheless, it will be frustrating to learn that for an extra few dollars you might have gotten an order of magnitude higher performance from a design.  As a scientist you are probably used to pursuing a problem to the best solution, but you shouldn't expect an engineer to pursue a problem past an adequate solution.  You have to understand and to communicate your goals. 

\subsection{Your People Skills are Crucial}

Your team is composed of humans with emotions and expectations. They will be more productive and communicate better if they enjoy working with you, respect you, and understand that you are invested in their success and happiness.  A wide range of styles may be successful, but insincerity is poisonous.  Learning about your team members interests builds bridges and exposes you to new areas that you might enjoy.

\section{Understand the Technology} \label{sec:understand}

\subsection{Ask Your Team a Lot of Questions}

It is not rude to closely examine and discuss your team's work.  Most designs and processes aren't born perfect.  Most are a little too complex or neglect a constraint that might later cause problems.  The process of explaining the design to another person frequently spurs a thought process and comments that lead to improvement.  If members of your team are reluctant to engage in this way you have identified a problem that needs correction.  If the design is already close to perfection, you should express your appreciation.

\subsection{Pay Attention to Details}

Paying attention to details motivates asking a lot of questions and thinking carefully about the design.  Important details include improving serviceability and averting safety issues.  These details may not change the design fundamentally but may save time and money during the instrument life cycle.  You will need to open your instrument for inspection, cleaning, and service over its lifetime, so consider how to ease access. Attention to eliminating assembly obstacles is another important serviceability concern. Team members will be upset if they cut themselves on thoughtlessly exposed sharp sheet metal or protruding hardware.  The discipline to consider serviceability does not come naturally to many engineers.  

I read recently about a high-end sports car with a ridiculously complex lubrication system that requires \$20,000 oil changes.  You will be remembered, not fondly, if your designs require heroic maintenance.  I recall many years ago looking at an instrument that we had just assembled and realizing that to access the interior we would have to repeat the entire assembly process in reverse.  We had included no access hatches.

\subsection{The Role of New Technologies}

There is a big difference between working at the cutting edge and at the bleeding edge. Very few instrument groups have the human or financial resources to really develop new technologies.  I have witnessed a number of painful failures resulting from underestimating the difficulty of technical development.  My goal has been to apply recently developed technology where warranted, but to watch carefully for the onset of the quicksand.  The promise of an advanced detector is the most frequent culprit in enticing instrument builders into the quicksand.  Proceed with extra caution when your detector system is ``almost there".

How does one stay in touch with new technologies to determine when they have crossed the line to cutting edge?  Internet search tools are increasingly powerful and allow you monitor new developments in technologies that are mature enough to have entered the literature.  Talking with a variety of colleagues actively engaged in building instruments can alert you to new technical developments that you will want to watch.  You may not initially have a wide range of contacts, so you should develop them by attending large instrument conferences (SPIE conferences are one example) or visiting institutions developing instruments and meeting people.  If you want to lead instrument development you need to learn how to approach people and engage them in conversation.  Usually you will get a friendly response, but if not move on.

When I began to be interested in ground-based instruments, I didn't have a budget for travel to international meetings.  I read every recent instrument conference proceedings that I could find in our library (well before NASA's Astronomical Data Service).  I learned as much or more than I might have learned by attending the conferences, but of course made no contacts.  I did learn whom I wanted to meet.

\clearpage

\section{Expertise}

Building instruments is an exercise in applied physics and requires specialized knowledge. I touch on several areas that the leader of an instrument project should aim to command.  The relative importance of subfields depends on what type of instruments you work with.  My experience has been mainly with optical and near-infrared imagers and spectrographs so that colors my discussion below, particularly in the order that I approach the topics.

\subsection{Optical Design}

Despite their importance, courses in optics have been removed from many, if not most, physics and astronomy curricula. If you were trained as a physicist or astronomer, you will probably have to educate yourself on optics. Although a knowledge of physical optics is very useful, a working knowledge of geometrical optics is essential.  A wealth of books is available and I recommend a few in the suggested reading section below.  You may find useful courses in university engineering departments unless you are near optics centers like the University of Rochester or the University of Arizona.

A good way to see if your understanding of optics is progressing is to choose a well-documented instrument that interests you and to try to understand its optical design.  If you can get access to an optical design code like ZEMAX, model the system and see how light propagates through the system.

One of the best ways to learn about optics is to work on designs with an expert.  When I began to work on the MMT conversion I had learned that Harland Epps was the go-to optical designer for spectrograph and telescope optics.  I wrote to Harland and asked him if he was interested in working on the MMT conversion and its instruments.  He fortunately agreed and explained his consulting arrangements.  My institution made the arrangements and I began one of the most productive collaborations of my career.  By visiting Harland and discussing the designs, I received a master class in optics.  After Harland retired, I was forced to develop optical designs on my own, but I built on what he taught me.

\subsection{Structures}

The backbone of your instrument is a structure that holds its components in place under changing environmental conditions, including gravity and external temperature.  It is important to understand the design of efficient structures.  An efficient structure achieves high stiffness (low flexure) without excessive mass, while allowing access to the internals of your instrument. Your background in physics taught you the basics of statics and dynamics.  Your engineering team will use finite element codes to analyze the performance of your instrument structure, but look carefully at the structure to verify that it began with a thoughtful concept.  Unfortunately, the structure of many instruments is an afterthought, rather than the result of a careful initial plan.  It is generally difficult to stiffen instruments at the end of the design cycle.

\subsection{Mechanisms and Motion Control}

Some instruments have no moving parts, but most have a few.  Linear or rotary motion stages are the most common type of mechanism that you will encounter.  If there are mechanisms in your instrument, you need to know something about motors, motor controllers, gearboxes, bearings, and guide rails.  Brushless DC motors, strain wave gearboxes, and recirculating ball guide rails are quality building blocks for a precision motion system.  Ask your engineering team about their component choices and make sure that you understand these choices.  Frequently, the cost difference between "probably ok" and "rock solid" is a small factor in the instrument budget.

\subsection{Detectors}
Mastering detector development typically involves cryogenics, vacuum technology, and very specialized electronics.  Detectors are frequently intolerant of static and misapplication of key signals and voltages.  This complexity and sensitivity, coupled with part cost and long delivery schedules, argues for special treatment. Over most of my career, I have had the good fortune to work with experienced detector scientists including John Geary and Peter Doherty.  Many institutions do not have the resources for a staff detector scientist.  Even if your institution does have a staff detector scientist, working with consulting firms that provide dewars, readout electronics, or complete integrated systems can make sense technically and financially.

\subsection{Electronics}

Instrument electronics generally include control electronics to configure instrument hardware for a particular observation and to control the detector to obtain the desired exposure. Electronics design and development evolve rapidly, but fortunately, highly developed and robust commercial products suitable for ground-based instruments are available.   Frequently, the major electronics design tasks are choosing appropriate components, packaging design, thermal management and cabling.  If a completely custom electronics package is proposed to you, ask why this approach is necessary.  If you are working on a space-based instrument, careful custom designs using space-agency approved devices are to be expected.

\subsection{Software}

Instrument software typically falls into three categories: instrument control, data reduction, and observation planning.  Instrument control software provides a user interface, communications with and control of configurable components, data taking and storage, communications with the host telescope, instrument health monitoring, and operational status logging.  Data reduction software reduces raw instrument data (often images), removes the instrumental signature, and produces calibrated data ready for scientific analysis.

The different categories of software require different knowledge and skills, and are frequently written by different groups. Because the instrument will not operate without instrument control software, it is often completed well before the data reduction software.  Sometimes, the data reduction software lags the nominal instrument commissioning. This delay is a problem because until reduced data flows the instrument cannot be judged working or commissioned.  The instrument has no scientific impact until reduced data are published.

The observation planning software typically also lags the instrument control software.  An exposure time calculator is a common component of observation planning software that predicts instrument performance for a variety of instrument configurations and observing conditions.  Moving the development of the exposure time calculator to early in the project can be very helpful in understanding and setting instrument requirements.  Confronting what the data might yield as early as possible keeps the focus on your scientific objectives.

\subsection{Vendors}

The interaction with vendors is simpler if you purchase only catalog items for your instrument.  For catalog items, the main worry is that delivery times may be inaccurate and/or quite long for low volume items.  When you ask for quotes on custom items there are several useful approaches.  Before requesting a detailed price and schedule quote, make sure that you have clearly specified your requirements.  Vendors dislike burning unpaid time generating repeated quotes on moving targets.  Don't wear out your welcome before you get started. Do get advice from others in the field about reliable vendors producing particular custom parts and the best engineering contacts at those vendors. If you ask for a firm, fixed price quote the price shouldn't evolve (again, try not to change your requirements!) but you may find that the schedule drifts.  You typically have little recourse once you have committed, so probe the schedule carefully before accepting the quote.  

When contracting with a vendor for expensive custom items you will need to prepare a specification document and detailed drawings.  You will have little recourse for problems that arise in areas that are not documented in the contract.  However, once you have provided the appropriate documentation don't assume that the vendor will read all the material carefully.  The more complex the documentation, the more likely it is that the vendor will miss important information by just skimming your documents.  You need to discuss the key points directly, preferably with the personnel at the vendor who will actually do the work.  Your contractual protection doesn't prevent schedule delays if a part has to be remade or reworked. 

Some years ago during the procurement of the optics for the MMT Wide-Field Corrector, the vendor was having difficulty achieving a good figure on one of the surfaces and argued that the surface met our specifications.  I pointed out that on page 26 of my elegant 40-page specification this situation was clearly addressed. The vendor team admitted that they hadn't consulted the document since their proposal.  I remember being surprised, even shocked, by the admission.

\section{The Life Cycle of an Instrument}

\subsection{Developing the Instrument Concept}

An instrument concept is a balancing act between the financially and physically achievable and the exciting scientific potential.  However, low scientific impact is not counterbalanced by excellent hardware.  Some poorly designed instruments manage to obtain transformational data and some elegantly executed instruments never have much impact.  If you are building an instrument dedicated to your own science, you can be assured that the data will be transformational!  But many times the user group is larger, with diverse interests.  Having the scientific background to judge the potential scientific impact of your instrument is important.  If you are early in your career it is important to broadly understand what is going on in your field.  Attend as many scientific talks in your field as possible.  You should not believe that you are too busy to learn about the science.

One of the most powerful lessons in science outweighing elegant technology was brought home to me when I found an image of the state-of-the-art (circa 1930) Mt. Wilson 100 inch spectrograph.  Although the fast spectrograph camera was advanced for the time, the main structure of the instrument is a text-book example of how not to do it.  The structure is four parallel tubes connecting the mounting flange and the main optics assembly, innocent of any attempt at a truss or triangular stiffening.  Hubble and Humason used the spectrograph in this poorly-designed structure to discover the expansion of the Universe.  Sadly, none of my spectrographs have had this degree of impact.

Developing an instrument concept requires a quantitative understanding of the key observations the instrument must make to be successful and how these translate into the instrument design.  This sounds like the basis of systems engineering and it is. 

\subsection{Funding and Schedule}

Unless you have already built a very similar instrument with the same team or are buying it from a catalog your initial thoughts on budget and schedule are probably grossly wrong.  You can guess the sign of the likely error.  The most expensive part of your project is team salaries, so getting the schedule wrong usually has significant cost consequences.

How can reliable cost estimates be obtained?  The most helpful tool is experience, so if you are starting out, talk with team leaders of other instruments and ask for the final costs of their instruments.  You may be lucky enough to have a very experienced project manager for your project or available to consult.  An unreliable cost estimate is easy to obtain; one common approach is to add up the cost of all of the instrument parts that you can think of and ask an engineer how long the design will take.

Once you have prepared a cost estimate you have the challenge of obtaining funding.  The process depends on the funding source, but it will usually be the science, not the instrument that attracts the funding.  A plausible approach and a competent team for building the instrument are required, but are usually of secondary importance at the funding stage.  The incentive for getting the design, schedule, and cost correct is that serious failures will be fatal to the project if the funding is fixed. Some of the least pleasant discussions in my instrument building career have been about cost growth.  If my institution had decided not to cover the increased costs, the discussions would have been far more unpleasant.

If only modest funding is available at the beginning of a project, these resources can be wisely invested to resolve technical concerns that would seriously compromise the instrument, refining the optical design and formulating a clean structural design. At this stage, be careful undertaking work that really requires a full engineering team and will need to be repeated when more resources are available.

\subsection{The Design Process}

Key elements of the design process are described in Section \ref{sec:understand}.  Holding frequent design progress discussions on key subsystems with the responsible engineers will help you identify where the design is incomplete or flawed.  You should understand the designs and the thinking behind them.  The engineers may have more design experience than you, but the process of discussing designs frequently leads to improvements. You will also have more at stake than anyone else in the design process. Larger reviews with external reviewers are helpful, but these high level reviews can miss important details. Nonetheless, the process of preparing for an external review is valuable.  As you prepare the review documents you should ask yourself if you have a convincing case for all aspects of the design.  If you have questions, you should answer them before the review.  If reviews identify problem areas, you will need to return to the design to resolve them.  Being closely involved in the design process from the beginning helps avoid later issues.

\subsection{Assembly, Integration and Test}

An efficient assembly, integration and test phase requires considerable planning.  Will you have the parts, electronics and software available to allow quality control, subsystem assembly, and subsystem testing?  Frequently, additional fixtures are needed to assemble an instrument and test its sub-assemblies before their integration into a complete instrument.  These fixtures may be small to hold tiny parts during assembly, or large to lift heavy optics assemblies safely, but each requires design time and should be accounted for in your budget and schedule.

When parts arrive and are assembled into instrument subsystems you will want to have electronics and software ready to begin testing.  The best time to identify and correct problems with your instrument is at the subsystem level before integration into a larger structure.  Watch the subsystem work.  Can you see any unanticipated behavior or hear odd noises? Don't ignore anomalies - they may be be early warning signs of problems that should be addressed with redesign or adjustments.  Hardware will not fix itself - that's your job.  As painstaking as the test phase may seem, it is always easier to identify and fix problems in the lab before shipping the instrument to the telescope.

\subsection{When to Begin Commissioning}

You should only consider commissioning your instrument at the telescope when you have completed an exhaustive testing program in the lab. Every part of the instrument should be functioning as designed.  All of your control software with user interfaces should be complete and communications with the telescope should be tested.  Very few people are as effective in a distant location in the middle of night as they are in normal daylight hours in the lab.

\subsection{Support Your Instrument's Success After It is in Service}

It's quite possible that problems crop up after commissioning is formally complete.  Pay attention to reports of unexpected instrument behavior and address problems promptly.  Software usually requires several iterations to become fully reliable and fully functional.  You need to honestly confront what you did right or wrong if you are going to become a better instrumentalist.  If you build ground-based instruments you can make changes and improve instrument operation after commissioning, and you should use the opportunity.

\section{Write the Instrument Paper}

Do not allow more than a few months to pass before you begin to write an instrument paper that documents instrument features and performance.  You will find this a useful resource years into operations when you might forget details.  You should plan on writing an instrument hardware manual with details of all of the parts that went into the instrument and service procedures.  An observer's manual describing instrument operations, features and software interfaces from an observer's perspective is also essential.

Margaret Geller played an important role in my career, motivating and supporting my early  instruments. I learned many things from Margaret, among them that our great insights do not exist for most of the world unless they are published.  She reminded me more than once that it was time to write the instrument paper.  These papers are important for the instrument building community and allow the instrument builders to take stock and to share their insights.

\section{Useful Technical Skills}

Most students completing an undergraduate major in the physical sciences or engineering will typically have the necessary software skills to work with instruments.  Acquiring a background in hands-on techniques is more typically where additional training is required.

Although it is unnecessary to be a skilled machinist, knowledge of how parts are made in a machine shop is very helpful in the design process to avoid unnecessary production difficulties.  Machinists are usually happy to critique your parts and suggest improvements, but gaining a basic understanding of how machine tools work is a good place to start.

Some proficiency in CAD software is useful if you are motivated and have an expert to consult when the CAD interface becomes baffling.  You can get access to student or ``maker'' versions of common CAD software at a reasonable price.  With CAD software you are empowered to rapidly create 3D printed parts to test ideas.  In most cases, I recommend leaving the heavy CAD work to engineers and designers with years of experience.

Electronics evolve rapidly, and many instruments rely heavily on commercial parts, but having some background in electronics will be helpful.  There are many on-line resources, but the classic reference for astronomers and physicists is \cite{2015Horowitz}.  Learning how to solder and assemble basic circuits and cables will be useful in the field.  Putting together a simple kit or two will help you gain those skills.  So much of modern electronics operate in the digital world that analog design skills are becoming rarer.  A scientist with a background in physics and some knowledge of electronics can sometimes contribute usefully in that environment.

Learning how to assemble optical systems and aligning optics cannot easily be learned from a book.  Volunteering to help an experienced colleague assemble and align their instrument is the most practical way to learn the basics.

\section{Useful References}

\cite{1999Schroeder} and \cite{1958Strong} are two of my favorite books on optics. \cite{2009bsa..book.....M} is a useful compendium of experimental methods that is based on the classic \cite{1938Strong}. \cite{2008McLean} and \cite{2001Janesick} provide an introduction to astronomical detectors.  \cite{2020Palmer} is a classic introduction to diffraction gratings.  \cite{2004SPIE.5492..553F} and \cite{2006SPIE.6269E..5TF} describe the basic concepts of RTV-based and flexure-based lens and mirror mounts.

\begin{acknowledgments}
I thank all of those who helped me learn and tolerated my mistakes.  I thank the anonymous referee for thoughtful comments that improved the manuscript.
\end{acknowledgments}

\bibliography{WhatIWish}{}
\bibliographystyle{aasjournal}

\end{document}